\documentstyle[12pt,epsf]{article}
\newcommand{\wh}{\widehat}
\newcommand{\eps}{\epsilon}
\newcommand{\vn}{\wh n}
\newcommand{\MM}{{\cal M}}
\newcommand{\LL}{{\cal L}}

\newcommand{\be}{\begin{equation}}
\newcommand{\ee}{\end{equation}}
\newcommand{\ben}{\begin{eqnarray}\displaystyle}
\newcommand{\een}{\end{eqnarray}}
\newcommand{\refb}[1]{(\ref{#1})}

\begin{document}

{}~ \hfill\vbox{\hbox{hep-th/9711130}\hbox{MRI-PHY/P971131}
}\break

\vskip 3.5cm

\centerline{\large \bf String Network}

\vskip 1cm

\centerline{\large \rm Ashoke Sen
\footnote{E-mail: sen@mri.ernet.in}}

\vspace*{1.5ex}

\centerline{\large \it Mehta Research Institute of Mathematics}
 \centerline{\large \it and Mathematical Physics}

\centerline{\large \it  Chhatnag Road, Jhoosi,
Allahabad 221506, INDIA}

\vspace*{4.5ex}

\centerline {\bf Abstract}

Type IIB string theory admits a BPS configuration in which three
strings (of different type) meet at a point. Using this three
string configuration we construct a string network and study its
properties. In particular we prove supersymmetry of this
configuration. We also consider string lattices, which can be
used to construct BPS states in toroidally compactified string
theory.

\vfill \eject

\baselineskip=18pt

Type IIB string theory in ten dimensions is known to have a
stable configuration in which three strings of different type
meet\cite{SCHWARZ,AHA,ZWI,SUKE}. If the three strings are of type
$(p_i,q_i)$ ($1\le i\le 3$)\cite{STHA,SCHOR,WITT}
then charge conservation requires
\be \label{e0}
\sum_{i=1}^3 p_i =\sum_{i=1}^3 q_i = 0.
\ee
\begin{figure}[!ht]
\begin{center}
\leavevmode
\epsfbox{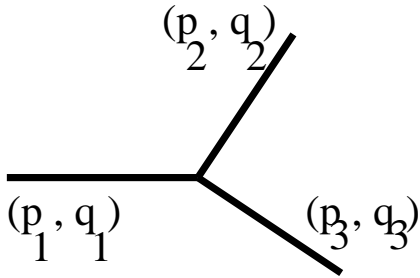}
\end{center}
\caption[]{\small Three string junction.}
\end{figure}
\noindent The configuration is shown in Fig.1. 
The angles between different
strings are adjusted such that the net force on the vertex due to
the tensions between different strings cancel\cite{SCHWARZ}. 
If $T_{p,q}$
denotes the tension of a $(p,q)$ string and $\wh n_i$ denotes the
direction of the $i$th string meeting at the vertex, then
we must have
\begin{figure}[!ht]
\begin{center}
\leavevmode
\epsfbox{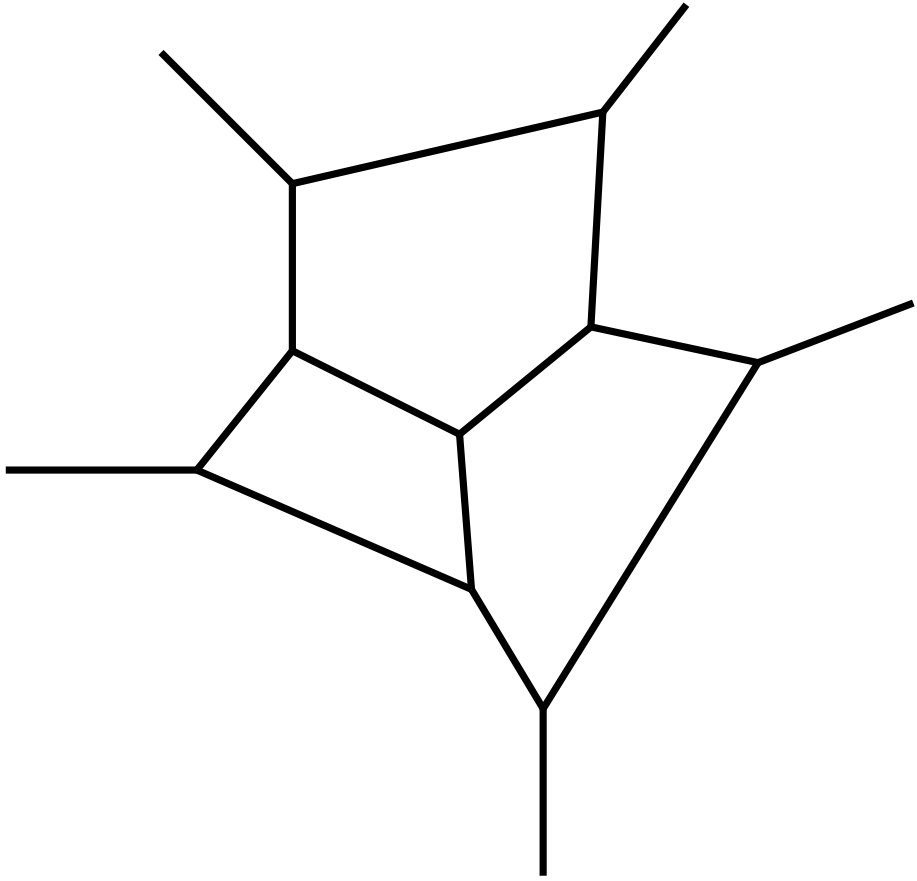}
\end{center}
\caption[]{\small String network}
\end{figure}
\be \label{e1}
\sum_{i=1}^3 T_{p_i,q_i} \wh n_i =0\, .
\ee
The BPS nature of this configuration was proved recently in
\cite{SUKE} (See refs.\cite{CM,GI,PE,TH,HA} for related work). 
Given such a configuration, one can also construct
string network by joining many of these vertices
together\cite{SCHWARZ}, as shown in Fig.2, with eqs.\refb{e0}, \refb{e1}
satisfied at each vertex. In this paper we shall discuss some of
the properties of such a network, including their supersymmetry
properties. For simplicity, we shall restrict our analysis to
planar network where all the strings lie in a single plane.

\noindent{\bf String Network}: Eq.\refb{e0} guarantees charge
conservation at each vertex.
Let us now examine the consequence of applying eq.\refb{e1} at
each vertex, with $\wh n_i$ now representing a two dimensional
unit vector. First, we recall from \cite{SCHOR} that 
\be \label{e2}
T_{p,q} = {1\over \sqrt\tau_2} |p + q\tau|\, ,
\ee
where $\tau=\tau_1 + i\tau_2$ denotes the 
axion-dilaton modulus of type
IIB. Let us denote by $\phi(p,q,\tau)$ the argument of $(p+q\tau)$:
\be \label{e3}
(p+q\tau) = |p+q\tau| e^{i\phi(p,q,\tau)}\, .
\ee
{}From \refb{e0}, \refb{e2} and \refb{e3} we see that at each
vertex,
\be \label{e6}
\sum_{i=1}^3 T_{p_i,q_i} e^{i\phi(p_i,q_i, \tau)} = 0\, .
\ee
This shows that eq.\refb{e1} can be automatically satisfied if we
choose:
\be \label{e4}
\wh n_i = \big( \cos\phi(p_i,q_i, \tau), \sin \phi(p_i, q_i,
\tau)\big) \, .
\ee
(This corresponds to orienting a $(p,q)$ string along the vector
$(p+q\tau)$ in the complex plane.)
\refb{e4} must be satisfied by all links in the network. From
this we see that the planar network has the beautiful property
that the orientation of a given link is determined solely by the
charges $(p,q)$ carried by the link, and not its location in the 
network. Thus for example, two links in the network, each
carrying a (2,3) string, will always have the same orientation,
irrespective of however far apart they are in the network.
Conversely, we see that as long as the network has the property
that any $(p,q)$ type string is oriented along the  vector
$(p+q\tau)$ in the complex plane, the force balance condition at
any junction is automatically satisfied.

\noindent{\bf Supersymmetry}: We shall now argue that this
network is invariant under one fourth of the space-time
supersymmetry of type IIB string theory. We proceed as follows.
Let us consider a $(p,q)$ string stretched along the 9th
direction. Let $\eps_L$ and $\eps_R$ be the two real supersymmetry
transformation parameters
of type IIB string theory, associated with the left
and the right moving sector of the world-sheet of the fundamental
string. The $(p,q)$ string configuration described above is
invariant under half of the supersymmetry of type IIB string
theory, generated by $(\eps_L,\eps_R)$ satisfying\footnote{Here
we are only considering asymptotic $\eps_L$ and $\eps_R$, and not
taking into account the space dependence that will be induced due
the classical background fields produced by the string.}
\be \label{e7}
\eps_L + i\eps_R = e^{i\phi(p,q,\tau)} \Gamma_1 \cdots \Gamma_8
(\eps_L - i\eps_R)\, ,
\ee
where $\Gamma_\mu$ are the ten dimensional gamma matrices. This
equation can be proved by first noting that for the (1,0) string
this gives:
\be \label{e9}
\eps_L = \Gamma_1\cdots \Gamma_8 \eps_L, \qquad
\eps_R = -\Gamma_1\cdots \Gamma_8 \eps_R\, ,
\ee
which is the correct formula.\footnote{Note that type IIB string
theory has left-right exchange symmetry on the world-sheet under
which $\eps_L\leftrightarrow\eps_R$. However, the presence of a
fundamental string of a given orientation breaks this symmetry, 
thereby giving rise to different conditions on $\eps_L$ and
$\eps_R$.} Furthermore, eq.\refb{e7} is invariant under the
SL(2,Z) transformation:
\be \label{e10}
\tau \to {a\tau+b\over c\tau+d}\, , \qquad \pmatrix{p\cr q}
\to \pmatrix{a & -b \cr -c & d} \pmatrix{p\cr q}\, ,
\ee
\be \label{e11}
(\eps_L-i\eps_R) \to \exp\Big({i\over 2} \arg(c\tau+d)\Big)
(\eps_L -i\eps_R)\, .
\ee
The transformation \refb{e11} of $(\eps_L-i\eps_R)$ can be
derived by repeating the arguments of
\cite{ORTIN} for type IIB theory. This shows that eq.\refb{e7} can be
derived by making an SL(2,Z) transformation of the corresponding
equations \refb{e9} for the fundamental string.

Let us now use this formula to test the supersymmetry of the
string network. Let us consider the $i$th link carrying charge
$(p_i,q_i)$ and oriented at an angle $\phi(p_i,q_i,\tau)$
relative to the 9-axis in the 8-9 plane. Eq.\refb{e7} is then
modified to:
\be \label{e12}
\eps_L + i\eps_R = e^{i\phi(p_i,q_i,\tau)} \Gamma_1 \cdots \Gamma_7
\big(\Gamma_8 \cos\phi(p_i,q_i,\tau) +
\Gamma_9\sin\phi(p_i,q_i,\tau)\big)
(\eps_L - i\eps_R)\, .
\ee
This equation can be satisfied {\it for all i} by requiring that
\ben \label{e13}
&& \eps_L = \Gamma_1\cdots \Gamma_8 \eps_L, \qquad
\eps_R = -\Gamma_1\cdots \Gamma_8 \eps_R\, , \nonumber \\
&&\eps_L = \Gamma_1 \cdots \Gamma_7 \Gamma_9 \eps_R\, .
\een
This reduces the supersymmetry to one fourth of the original
amount. Since eqs.\refb{e13} are independent of the composition
of the network, but dependent only on its orientation, we see
that a configuration of two or more parallel networks is 
invariant under the same supersymmetries even if the different
networks have different compositions.
Therefore we expect such
configurations to form a stable system.

\begin{figure}[!ht]
\begin{center}
\leavevmode
\epsfbox{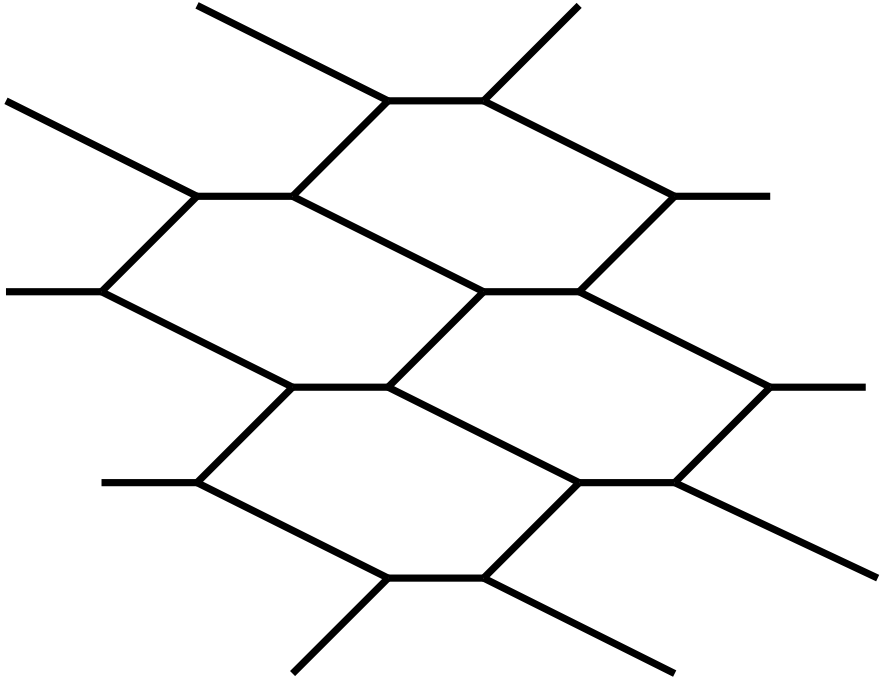}
\end{center}
\caption[]{\small String lattice}
\end{figure}
\noindent{\bf String Lattice}: By taking a periodic network we
can construct `string lattices'. An example has been shown in
Fig.3. Here the links of three different orientations correspond
to strings carrying three different kinds of charges. Note that
although the orientations of the links are fixed by the
charges they carry, their lengths are arbitrary. Thus the lattice
displayed in Fig.3 is characterized by the three length
parameters $l_1$, $l_2$ and $l_3$, $-$ the lengths of the three
different kinds of links in the lattice. These three parameters
determine the length of the two independent basis vectors of the
lattice and the angle between these basis vectors.

This configuration can also be interpreted as 
a BPS state of type IIB string
compactified on a two dimensional torus, with periodicity
matching the periodicity of the lattice. Thus $l_1$, $l_2$ and
$l_3$ determine the shape and size of the torus. Conversely, for
a given torus, one needs to fix $l_1$, $l_2$ and $l_3$
appropriately so that the network fits on the torus. 

The mass of the BPS state is given by the total mass of all the
strings that lie inside a unit cell.
For the lattice shown in Fig.3, the unit cell can be identified
as the parallelogram obtained by joining the centers of four
adjacent hexagons.
If $(p_i,q_i)$ denote the charges carried by the $i$th type of
link in the lattice, then the mass of the BPS state represented
by this lattice is given by
\be \label{e14}
\sum_{i=1}^3 l_i T(p_i,q_i)\, .
\ee
One can bring this formula in a more conventional form as
follows. If $\vn_1$, $\vn_2$ and $\vn_3$ denote the unit vectors
along the three links meeting at a vertex, then the two
independent basis vectors of the lattice can be taken as:
\be \label{e15}
\vec a = l_1 \vn_1 - l_3 \vn_3, \qquad \vec b= l_2 \vn_2 - l_3
\vn_3\, .
\ee
The area $A$ and the modular parameter
$\lambda=\lambda_1+i\lambda_2$ of the torus are then given by
\be \label{e16}
A = |\vec a \times \vec b|\, ,
\ee
\be \label{e17}
\lambda_1 = {\vec a\cdot \vec b}/\vec a^2, \qquad \lambda_2
= |\vec a\times \vec b|/\vec a^2\, .
\ee
It is a straighforward (although somewhat tedious) exercise to
see that in terms of $A$ and $\lambda$, the mass formula
\refb{e14} can be rewritten as
\be \label{e18}
m^2 = A \pmatrix{p_1 & q_1 & p_2 & q_2} (M \pm L) \pmatrix{p_1
\cr q_1\cr p_2 \cr q_2}\, ,
\ee
where,
\be \label{e19}
M = {1\over \lambda_2} \pmatrix{ \MM & \lambda_1 \MM \cr
\lambda_1 \MM & |\lambda|^2 \MM}\, , \qquad L = \pmatrix{0 & \LL
\cr -\LL & 0}\, ,
\ee
\be \label{e20}
\MM = {1\over \tau_2} \pmatrix{1 & \tau_1\cr \tau_1 & |\tau|^2}\, ,
\qquad  \LL = \pmatrix{ 0 & 1\cr -1 & 0}\, .
\ee
For given $(p_i,q_i)$,
the sign in front of $L$ in eq.\refb{e18} is chosen such that
the contribution from this term to $m^2$ is positive. 

This is the $SL(2,Z)_S\times SL(2,Z)_U$ invariant BPS formula 
for type IIB string theory on $T^2$.
Note that in this case the full U-duality group is $SL(3,Z)\times
SL(2,Z)$, but the mass formula \refb{e18} does not display this
symmetry manifestly, since we have considered states carrying no
momenta along the internal directions of the torus, and also have
set the anti-symmetric tensor field background to zero. Also note
that using an appropriate SL(3,Z) transformation we can convert
the charges $(q_1,q_2)$ associated with Ramond-Ramond tensor field
to 
internal momenta on the torus. Thus under this duality
transformation the BPS states represented by the string lattice
get mapped to elementary string states. It will be interesting to
compare the degeneracy of states considered here with that of the
elementary string states.

This finishes our short note on string network. At present the
utility of the string network, besides describing BPS states in
toroidally compactified type IIB string theory, is not clear.
However, in future a manifestly SL(2,Z) invariant
non-perturbative formulation of string theory may be
made possible by
regarding the string network, instead of string loops, as
fundamental objects. This would be similar in spirit to recent
developments in canonical quantum gravity, in which loops 
have been replaced by spin networks\cite{ROVELLI}.

Note added: After writing this paper I became aware of
refs.\cite{ONE,TWO} where supersymmetric configurations of webs
of strings and five-branes have been discussed. Some related work
has also been reported in refs.\cite{KOL,SHIM,VAFA}.

\end{document}